\documentclass[reprint,aps,prl,superscriptaddress]{revtex4-2}
\usepackage{graphicx}
\usepackage{amsmath,amssymb,bm}
\usepackage{hyperref}
\usepackage{xcolor}
\usepackage{soul}
\usepackage{cancel}
\usepackage[percent]{overpic}
\usepackage[caption=false]{subfig}
\usepackage[version=4]{mhchem}

\begin{document}

\title{A Machine Learning Model for the Chemistry of a Solvated Electron}

\author{Ruiqi Gao}
\affiliation{Department of Electrical and Computer Engineering, Princeton University, Princeton, New Jersey 08544, USA}

\author{Pinchen Xie}
\affiliation{Applied Mathematics and Computational Research Division,  Lawrence Berkeley National Laboratory, Berkeley, California 94720, USA}

\author{Roberto Car}
\affiliation{Department of Chemistry, Princeton University, Princeton, New Jersey 08544, USA}

\date{\today}

\begin{abstract}
In molecular simulations, machine-learning force fields can achieve ab initio accuracy at a lower cost but remain limited in the explicit modeling of electrons. In this work, we develop an electron-aware machine-learning force field, in which an excess electron of interest is modeled quantum mechanically, while the remaining short-range interactions and long-range Coulombic forces are machine-learned to reproduce a density functional theory calculation. We demonstrate the method on the solvated electron in bulk water and its reaction with a hydronium ion. We identify a proton transfer mechanism by which the excess proton recombines with the electron. We determine the forward reaction rates between 350~K and 450~K from first-passage survival functions, which yield an Arrhenius relationship with an activation energy of 3.2~kcal$\cdot$mol$^{-1}$, in good agreement with experiment. From an enhanced sampling simulation, we determine the equilibrium constant, and thus the reaction free energy, which is also consistent with experimental measurements.
\end{abstract}

\maketitle


In recent years, machine-learning (ML) force fields have been developed
to learn from non-empirical quantum mechanical methods such as density functional theory (DFT),
enabling the simulation of large systems and long timescales with high accuracy \cite{behler2007generalized, schutt2017schnet, zhang2018deep, batzner20223}.
These force fields represent the adiabatic potential energy $E$ as a function of atomic coordinates $\{\mathbf{R}_i\}$ using neural networks.
While successful in many situations, including reactive processes involving bond breaking, these models lack explicit electron-level resolution. 
Additional ML models have been developed to predict bonding-electron positions in insulators  \cite{zhang2020deep},
and to incorporate them into long-range Coulomb interactions \cite{zhang2022deep}.
However, these do not apply to an electron not uniquely associated with an atom.
The solvated electron is one notable example. Despite recent attempts with a plain ML force field \cite{lan2021simulating, lan2022temperature, gao2024enhanced} to model its dynamics,
a robust and extensible ML solution has yet to be established \footnote{See Appendix F for the limitations of these approaches.}.

The solvated electron is a localized excess electron in a solution, stabilized by solvent polarization \cite{schindewolf1968formation, herbert2017hydrated}. It is a widely occurring intermediate in radiation-induced chemical and biological processes \cite{herbert2017hydrated, choppin2013radiochemistry}.
In this work, we focus on the solvated electron in water, e$^-_{\mathrm{(aq)}}$,
and its reaction with a hydronium ion:
\begin{equation}\label{reaction}
\mathrm{e}^-_{\mathrm{(aq)}} + \mathrm{H}^+_{\mathrm{(aq)}} \rightarrow \mathrm{H}\cdot_{\mathrm{(aq)}}
\end{equation}
This produces a hydrogen atom in water. Reaction~(\ref{reaction}) proceeds on the Born-Oppenheimer (BO) ground-state surface and is arguably the simplest reaction of an electron in solution, but its detailed mechanism is still not clear \cite{herbert2017hydrated}.
A historically held view is that e$^-_{\mathrm{(aq)}}$ reacts via electron transfer \cite{HartAnbar1970},
but more recent DFT-based ab initio molecular dynamics (AIMD) \cite{car1985unified} for an excess electron in water clusters \cite{marsalek2010hydrogen} suggest a proton-transfer mechanism instead. Quantitative comparisons with the experiments were not possible because of the high computational cost of simulating such a process in bulk water.  

Separately, empirical pseudopotential models have been developed to represent e$^-_{\mathrm{(aq)}}$
\cite{schnitker1987electron, turi2002analytical, jacobson2010one},
in which a one-electron Hamiltonian for e$^-$ is coupled to empirical water models.
These methods can reproduce the cavity structure and key spectroscopic properties of e$^-_{\mathrm{(aq)}}$ \cite{herbert2017hydrated}, but lack the first-principles 
description of all electrons of DFT.
Additionally, they are tuned for an isolated e$^-_{\mathrm{(aq)}}$,
so their transferability to reactive environments involving an additional H$^+_{\mathrm{(aq)}}$ is uncertain.

In this work, we address the limitations of ML force fields using an electron-aware design. We introduce a hybrid method that treats the electron explicitly, reminiscent of one-electron pseudopotential approaches, and uses ML to describe short-range interactions and long-range Coulombic forces. Our simulations support a proton-transfer mechanism for the reaction~(\ref{reaction}) and yield reaction rates and equilibrium constants in good agreement with experiment. Central to our approach is the following ansatz for the total potential energy:
\begin{equation}\label{E}
E_{\mathrm{total}} = E_{\mathrm{se}} + E_{\mathrm{lr}} + E_{\mathrm{sr}}
\end{equation}
Here, $E_{\mathrm{se}}$ is the ground-state energy of a one-electron Schrödinger equation,
which includes the long-range interaction between the excess electron and the rest of the system, as well as an approximate kinetic energy of localization.
$E_{\mathrm{lr}}$ describes the long-range Coulomb interaction within the rest of the system, excluding the excess electron.
$E_{\mathrm{sr}}$ is the short-range energy, represented with a typical neural network potential.
In the following, we define these terms in detail and describe the physical motivations that underlie this decomposition.

\begin{figure}[t]
  \centering
  \includegraphics[width=\linewidth]{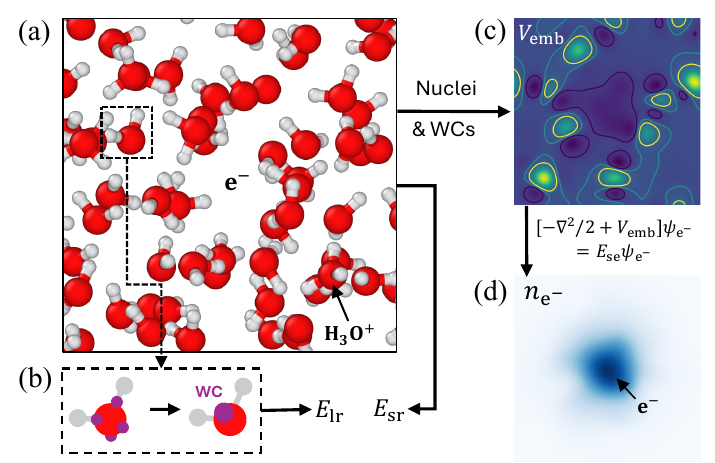}
  \caption{
  Illustration of our scheme for a periodic 64-H$_2$O cell with one H$^+$ and one e$^-$.
  (a) Sketch of the neighborhood of e$^-_{\mathrm{(aq)}}$.
  (b) MLWF centers (small purple dots) and WC (large purple dot) for a water molecule.
  (c) $V_{\mathrm{emb}}$ generated by $\rho_{\mathrm{emb}}$ (brighter=higher energy).
  (d) $n_{\mathrm{e}^-}$ from Eq.~(\ref{se}).
  (c,d) show the $xy$-plane slices centered at the MLWF center of e$^-$. }
  \label{fig:V_rho}
\end{figure}

We begin by formulating $E_{\mathrm{se}}$. Note that the energy function only takes the atomic coordinates as input, so our machine learning model does not directly see the solvated electron. To capture the physics, we need to identify the position where the electron localizes. We realize this by explicitly solving an auxiliary one-electron Schrödinger equation under the long-range potential generated by the environment, i.e., the system minus the excess electron, which consists of all atomic cores (nuclei plus core electrons) and all valence electrons of water. The atomic cores are located at the nuclear positions $\{\mathbf{R}_i\}$. The valence electrons can be approximated using Maximally Localized Wannier Function (MLWF) centers \cite{marzari2012maximally}.
Specifically, eight valence electrons are associated with each H$_2$O molecule,
and the corresponding MLWF centers can be identified and averaged to what is called a Wannier centroid (WC),
as shown in Fig.~\ref{fig:V_rho}(b).
The WC is located near the oxygen atom and carries a charge of $-8$e,
and its position $\mathbf{R}^{\mathrm{WC}}$ can be predicted by a neural network \cite{zhang2020deep} based on the atomic neighborhood of each oxygen atom.
Together, the total charge density of the environment $\rho_{\mathrm{emb}}$,
which contains all the atomic core and valence electron charges,
can be approximated as
\begin{equation}\label{rho}
\rho_{\mathrm{emb}}(\mathbf{r}) = \sum_{i} Z_i \rho_s(\mathbf{r} - \mathbf{R}_i) + \sum_{j} q_j \rho_s(\mathbf{r} - \mathbf{R}^{\mathrm{WC}}_j),
\end{equation}
with $Z_i$ being the atomic core charge, $q_j$ being the WC's charge, and $\rho_s$ being a spherical Gaussian-smoothed delta function.
This approximation is accurate up to the dipole order \cite{zhang2022deep}.
$\rho_{\mathrm{emb}}(\mathbf{r})$ generates a Coulomb \emph{embedding potential} $V_{\mathrm{emb}}(\mathbf{r}) = -\int d\mathbf{r}' \rho_{\mathrm{emb}}(\mathbf{r}')/|\mathbf{r} - \mathbf{r}'|$ for the electron,
and we can write the following auxiliary one-electron Schrödinger equation for the excess electron
\begin{equation}\label{se}
\left[-\tfrac{1}{2}\nabla^2 + V_{\mathrm{emb}}\right]\psi_{\mathrm{e}^-} = \varepsilon \psi_{\mathrm{e}^-}.
\end{equation}
The ground-state eigenvalue $\varepsilon$ from Eq.~(\ref{se}) defines $E_{\mathrm{se}}$.

We remark that we do not intend to recover the Kohn-Sham energy and wavefunction from Eq.~(\ref{se})\footnote{One difference is the nodal structure: the lowest eigenstate of Eq.~(\ref{se}) is nodeless, whereas the corresponding Kohn--Sham orbital is orthogonal to the occupied same-spin orbitals. Appendix~C compares other properties of the auxiliary and Kohn--Sham states.}.
We seek to locate the center-of-mass \footnote{For an orthorhombic cell, the periodic center of the electron density $n_{\mathrm e^-}$ along direction $\alpha$ is defined as
$c_{\mathrm e^-,\alpha}=(L_\alpha/2\pi)\operatorname{Im}\ln\!\left[\int_\Omega n_{\mathrm e^-}(\mathbf r)e^{2\pi i r_\alpha/L_\alpha}\,d\mathbf r\right]\pmod{L_\alpha}$.}
of the electron distribution and, subsequently, an interaction energy $E_{\mathrm{se}}$ that contains the Coulomb interaction up to dipole-order accuracy.
In Fig.~\ref{fig:V_rho}(a), we show a sample configuration of bulk water containing  e$^-_{\mathrm{(aq)}}$ and H$^+_{\mathrm{(aq)}}$.
The corresponding $V_{\mathrm{emb}}(\mathbf{r})$ in the cubic periodic cell is shown in Fig.~\ref{fig:V_rho}(c).
Solving Eq.~(\ref{se}) yields the density $n_{\mathrm{e}^-}=|\psi_{\mathrm{e}^-}|^{2}$ for the excess electron shown in Fig.~\ref{fig:V_rho}(d).
In Fig.~\ref{fig:V_rho}(c,d) the frames are centered at the MLWF center of e$^-_{\mathrm{(aq)}}$ obtained from DFT calculations and we can see that $n_{\mathrm{e}^-}(\mathbf{r})$ is localized around this MLWF center.
This also applies to the product state H$\cdot_{\mathrm{(aq)}}$, where the well of $V_{\mathrm{emb}}$ is centered on the excess proton. 
By construction, $V_{\mathrm{emb}}$ reproduces the exact tail of the electrostatic potential acting on the excess electron.  
This is sufficient for
the center-of-mass $\mathbf{c}_{\mathrm{e}^-}$ of $n_{\mathrm{e}^-}$ to approximate well the MLWF center of e$^-$ from DFT across the reactant, reactive, and product regimes, as shown in Appendix~B.
As long as $\mathbf{c}_{\mathrm{e}^-}$ is accurate, the long-range Coulomb interaction between e$^-$ and the rest of the system is correctly captured.
Any energy error due to the approximations is then a short-range effect and will be later absorbed into $E_{\mathrm{sr}}$ with a learnable neural network. 

Having established $E_{\mathrm{se}}$, we define $E_{\mathrm{lr}}$ and $E_{\mathrm{sr}}$ in~(\ref{E}).
First, $E_{\mathrm{lr}}$ is the long-range Coulomb energy
of the system excluding e$^-$, given by
\begin{equation}\label{lr}
E_{\mathrm{lr}} = \frac{1}{2}\int d\mathbf{r} d\mathbf{r}' \frac{\rho_{\mathrm{emb}}(\mathbf{r}) \rho_{\mathrm{emb}}(\mathbf{r}')}{|\mathbf{r} - \mathbf{r}'|}.
\end{equation}
This is, again, accurate up to dipole-dipole interactions.
The sum $E_{\mathrm{se}}+E_{\mathrm{lr}}$ constitutes the system's complete long-range electrostatics.
Next, the short-range energy satisfies $E_{\mathrm{sr}} = E_{\mathrm{total}} - E_{\mathrm{se}} - E_{\mathrm{lr}}$. $E_{\mathrm{sr}}$ has the typical form
$
    E_{\mathrm{sr}} = \sum_{i=1}^{N_{\mathrm{atoms}}} E_{\mathrm{NN}}(\{\mathbf{R}_j\}_{j \in \mathcal{N}(i)})
$
used in an ML force field,
where $\mathcal{N}(i)$ is the neighborhood of atom $i$ within a certain radius, and the local energy function $E_{\mathrm{NN}}$ is represented by a neural network. This is trained to capture all remaining short-range many-body interactions. This distinguishes our method from existing one-electron pseudopotential approaches~\cite{schnitker1987electron, turi2002analytical, jacobson2010one}, as our method is still a machine-learning method trained to faithfully reproduce the adiabatic potential energy surface of a given DFT functional. In addition, we do not target exact electronic energies or wavefunctions as these pseudopotential approaches do. Instead, we focus on reproducing the exact BO dynamics for large-scale simulations.

In summary, the full model of $E_{\mathrm{total}}$ is constructed on the basis of two neural networks: one predicts the WCs that underlie $E_{\mathrm{se}}$ and $E_{\mathrm{lr}}$, and another predicts $E_{\mathrm{sr}}$ directly. In practice, we first train the neural network that describes the WCs using DFT calculations of the WC positions \cite{zhang2020deep, marzari2012maximally}.
With this neural network, we can predict $\rho_{\mathrm{emb}}$ from a given atomic configuration, and $E_{\mathrm{se}}$ and $E_{\mathrm{lr}}$ are obtained from Eq.~(\ref{se}) and~(\ref{lr}), respectively. Next, using DFT data for total energy $E_{\mathrm{total}}$ and atomic forces $\mathbf{F}_{\mathrm{total}}$, the neural network representing $E_\mathrm{sr}$ is trained to predict $E_{\mathrm{total}} - E_{\mathrm{se}} - E_{\mathrm{lr}}$, as well as short-range forces $\mathbf{F}_{\mathrm{total}} - \mathbf{F}_{\mathrm{se}} - \mathbf{F}_{\mathrm{lr}} = \mathbf{F}_{\mathrm{total}} + \nabla_{\mathbf{R}} E_{\mathrm{se}} + \nabla_{\mathbf{R}} E_{\mathrm{lr}}$.
The derivative $\nabla_{\mathbf{R}} E_{\mathrm{se}}$ is obtained
via the Hellmann–Feynman theorem as
$\langle n_{e^-}, \nabla_{\mathbf{R}} V_{\mathrm{emb}}\rangle$.
The derivative $\nabla_{\mathbf{R}} V_{\mathrm{emb}}$, as well as $\nabla_{\mathbf{R}} E_{\mathrm{lr}}$, is obtained by directly differentiating the computational workflow,
which includes the neural-network prediction of the WCs. 
Energy is conserved in microcanonical simulations with our learned model.
Finally, we remark that WCs are determined by the local atomic environment~\cite{marzari2012maximally}, and $E_{\mathrm{sr}}$ is also expressed as a sum of local contributions. This makes our learned model extensible to larger systems. However, $\mathbf{c}_{\mathrm{e}^-}$ is a global quantity that depends on all atoms in the system to determine where it localizes. We achieve this dependency through $V_{\mathrm{emb}}$, a global function of all atoms. This overcomes the locality limitation of standard ML force fields in describing e$^-$. In addition, in our model the electrostatic interaction is treated explicitly, providing the physical driving force responsible for the stable localization of e$^{-}_{\mathrm{(aq)}}$.

\emph{Numerical implementation.} Our implementation, Deep~Potential~Long-Range~Quantum (DPLR-q), is based on the DP-MP neural network architecture implemented in the DeepMD-jax package~\cite{gao2024enhanced}. We use this architecture for both the WC and short-range models with default hyperparameter settings, including a 6.0~Å neighbor cutoff and one message-passing step.
The Schrödinger equation~(\ref{se}) is solved
on a real-space grid with an FFT evaluation of the kinetic energy operator.
We apply a Rayleigh--Ritz iteration on the span of the current iterate and its kinetic-preconditioned residual to efficiently obtain the ground state.
We evaluate $E_{\mathrm{lr}}$ on the same grid using the Ewald summation.
The resulting model maintains the linear-scaling and practical efficiency of standard DP-MP models \footnote{In the practical simulations of our system, the contribution of computational time from $E_{\mathrm{se}}$ is roughly equal to that from $E_{\mathrm{sr}} + E_{\mathrm{lr}}$}.

\emph{Model training and validation.}
We train our neural networks on the hybrid functional PBEh($\alpha=0.4$) plus the rVV10 van der Waals correction \cite{ambrosio2016structural}. This functional has been suggested to reproduce well the energy levels and other properties of e$^-_{\mathrm{(aq)}}$ \cite{ambrosio2017electronic, pizzochero2019picture}. Our initial dataset is generated using DFT molecular dynamics simulations of a periodic box containing 64 water molecules, an excess proton, and an excess electron. Details of DFT calculations and model training are given in Appendix~A.
Several iterations of active learning are performed following the protocol described in Ref.~\cite{zhang2020dp}. 
We test the robustness of our final model over 100~ns of combined simulation time from 350 to 450~K and do not observe any unphysical scenarios.
With a 90\%--10\% train--validation split of the dataset, the WC prediction achieves a root-mean-square error (RMSE) of $1.0\times 10^{-3}$~Å in distance,
and the short-range energy model has a 0.15~meV/atom energy RMSE and a 13~meV/Å force RMSE per Cartesian component. 
We then extend the system to 128 water molecules plus one proton and one electron at the same density
and conduct molecular dynamics simulations with our trained model. 
We observe the same reaction process as in the 64-H$_2$O cell.
Samples from these 128-H$_2$O trajectories form an additional validation set, for which we compute reference DFT data; on this set, our model achieves RMSEs at the same level as those above, confirming the extensibility of our model.

\emph{Reaction mechanism.}
Using the molecular dynamics driven by our $E_{\mathrm{total}}$ model, we accurately learn the reaction process between e$^{-}_{\mathrm{(aq)}}$ and H$^{+}_{\mathrm{(aq)}}$.
Before the reaction, H$^+_{\mathrm{(aq)}}$ diffuses in water through the Grotthuss hopping mechanism \cite{agmon1995grotthuss}.
Meanwhile, e$^-_{\mathrm{(aq)}}$ is localized in a cavity with a radius of $\sim\!\!2$~Å,
surrounded by 4--6 water molecules,
and diffuses through the exchange of water molecules in the solvation shell \cite{ambrosio2017electronic, lan2021simulating, gao2024enhanced}.
H$^+_{\mathrm{(aq)}}$ diffuses faster than e$^-_{\mathrm{(aq)}}$, and we observe a proton-transfer mechanism in the final reaction pathway.

\begin{figure}[t]
  \centering
  \includegraphics[width=0.95\columnwidth]{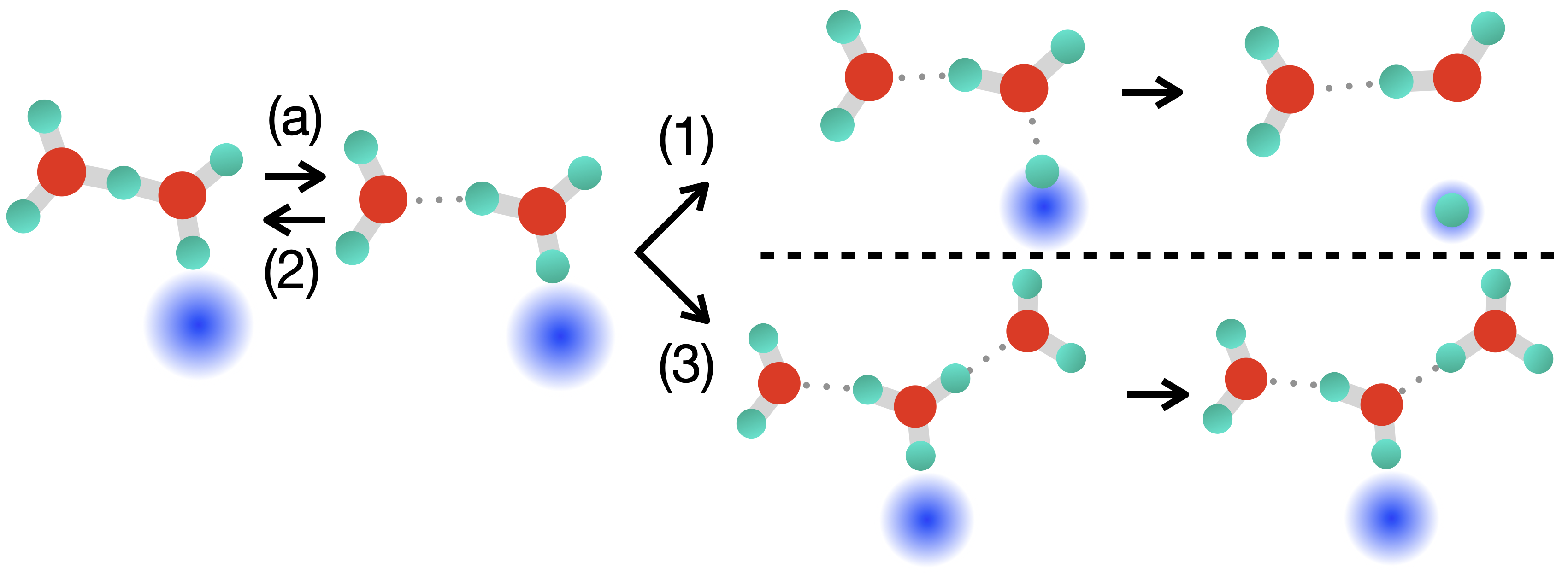}
  \caption{(a) The proton hops and forms an Eigen-like H$_3$O$^+$ beside the excess electron. Three outcomes follow: (1) reaction, (2) back-hop, and (3) forward relay. The excess electron (blue) is shown schematically; its spatial extent may exceed the rendering and can be non-isotropic during reaction~\cite{marsalek2010hydrogen}.}
  \label{fig:reaction_path}
\end{figure}

Fig.~\ref{fig:reaction_path} illustrates the mechanism we observe.
(a) The proton diffuses via Grotthuss hopping until it reaches the first coordination shell of e$^-_{\mathrm{(aq)}}$ from the outer side.
An H$_2$O molecule in the solvation shell of e$^-_{\mathrm{(aq)}}$ has one H pointing towards the electron;
with an additional proton, it forms an Eigen-like H$_3$O$^+$ structure.
Of the three hydrogens in H$_3$O$^+$, one points towards the electron, one points to the oxygen from which it comes, 
and the third points to another water molecule in the surrounding hydrogen-bond network.
From here, three pathways can occur.
(1) The hydrogen pointing to the electron breaks from H$_3$O$^+$ and forms the product H$\cdot_{\mathrm{(aq)}}$.
(2) The hydrogen pointing to the oxygen from which it comes hops back.
(3) The hydrogen relays forward to the other neighboring water molecule.
We observe all three pathways in our simulations, indicating complex and competing reaction dynamics.
Note that the reactants $e^-_{\mathrm{(aq)}}$ and H$^+_{\mathrm{(aq)}}$ are both strongly hydrated,
while the product H$\cdot_{\mathrm{(aq)}}$ is hydrophobic.
The reaction is driven by solvent polarization and involves substantial solvent reorganization, as is typical of charge-transfer processes in polar solvents \cite{chandler1998electron}, and is not a purely diffusion-limited reaction.

\emph{Reaction rate.}
Although the reaction is not diffusion-limited and has an energy barrier, the barrier is small enough that the process is accessible within our simulation timescale. A reaction is numerically identified when the distance between the excess proton \footnote{The excess proton is identified, following Ref.~\cite{marsalek2010hydrogen}, as the hydrogen remaining after assigning the two closest hydrogens to each oxygen.} and the excess electron center reaches below $0.3$~\AA. We can extract the reaction rate from the first-passage survival probability
by simulating an ensemble of independent trajectories.
To do this, we prepare an initial ensemble according to the quasi-stationary distribution (QSD) \cite{collet2012quasi},
i.e., the distribution of reactant configurations conditioned on not having reacted within a sufficiently long time.
This can be efficiently sampled by running a long trajectory with the following modification.
Whenever a reaction event is detected during the run,
we reset to a previous reactant configuration and randomize the velocities.
After a relaxation period following each reset, we start recording configurations until the next reaction event.
The collection of these surviving trajectory segments will converge to the QSD.
Let $k_2$ be the bimolecular rate constant for the reaction. The reactant concentration is $V^{-1}$ where $V$ is the volume of our simulation box.
Starting from the QSD, the survival probability $s$ is an exponential function of time~$\tau$, given by $s(\tau) = \exp(-\lambda \tau)$ with $\lambda = k_2 V^{-1}$.
We run $N_{\mathrm{traj}}$ independent trajectories starting from the QSD up to a right-censoring time $\tau_{\max}$ when at least half of the trajectories have reacted. 
We record the time-to-reaction $\tau_i$ and set $\tau_i = \tau_{\max}$ for a trajectory without a reaction. Using the maximum-likelihood estimator $\hat{\lambda} = (1 - \hat{s}(\tau_{\max}))/\langle \tau_i \rangle$, where $1 - \hat{s}(\tau_{\max})$ is the fraction of trajectories that have reacted by $\tau_{\max}$, we obtain $\lambda$ and thus $k_2$.

\begin{figure}[t]
  \centering
  \begin{overpic}[width=0.505\columnwidth]{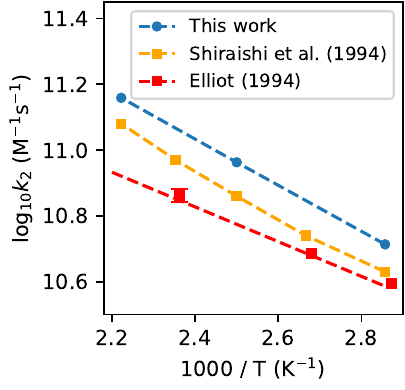}
    \put(28,23){\footnotesize (a)}
  \end{overpic}\hfill
  \begin{overpic}[width=0.485\columnwidth]{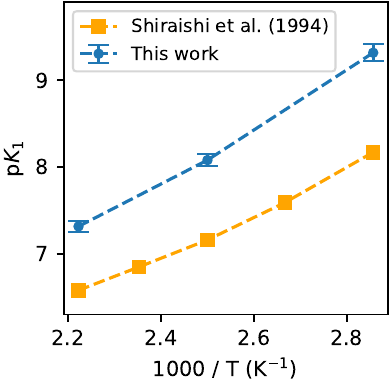}
    \put(88,23.5){\footnotesize (b)}
  \end{overpic}
  \caption{(a) Arrhenius plot for the predicted $\log_{10} k_2$ versus inverse temperature, compared with experimental data from Shiraishi et al.~\cite{shiraishi1994temperature} and Elliot~\cite{elliot1994rate}.
(b) van~'t~Hoff plot of the log of the equilibrium constant $K_1$ versus inverse temperature, compared with the experimental data of Shiraishi et al.~\cite{shiraishi1994temperature}.}

  \label{fig:rate}
\end{figure}

We perform this procedure at temperatures of 350, 400, and 450~K \footnote{As in Refs.~\cite{ambrosio2016structural, ambrosio2017electronic}, we simulate only at 350~K and above to enable a ``frank diffusive motion'' of water, since this DFT functional can show glassy behavior at lower temperatures.} 
using a system of 64, 128, 256, and 512 H$_2$O molecules with one proton and one electron.
For each condition, we sample from its QSD and run $N_{\mathrm{traj}}=2000$ independent trajectories.
The obtained $k_2$ values are extrapolated to the dilute limit
by a linear fit against the inverse box length,
and our result is shown in Fig.~\ref{fig:rate}(a).
An Arrhenius behavior ($k_2\propto e^{-E_a/RT}$) is observed, with the activation energy $E_a=3.2$~kcal$\cdot$mol$^{-1}$ (0.14~eV). This is in good agreement with experimental measurements \cite{shiraishi1994temperature, elliot1994rate}, where the results in Ref.~\cite{shiraishi1994temperature} at 400~K correspond to $E_a=3.4$~kcal$\cdot$mol$^{-1}$, and Ref.~\cite{elliot1994rate} reports $E_a=2.4$~kcal$\cdot$mol$^{-1}$.
The activation energy should be related to a reaction barrier in the process described in Fig.~\ref{fig:reaction_path}, since a diffusion-limited $k_2$ would be several times larger than the observed value \cite{shiraishi1994temperature, johnaelliot1990estimation}.

Still, our simulation displays a slight overestimation of the rate $k_2$ compared to experiments. We now discuss the sources of error in our simulation.
Firstly, the employed density functional approximation underestimates diffusion.
Calculation with our model at 350~K gives diffusion coefficients of $0.20 \pm 0.02$,
$0.30 \pm 0.03$, and $0.79 \pm 0.04$~Å$^2$/ps for H$_2$O, e$^-_{\mathrm{(aq)}}$, and H$^+_{\mathrm{(aq)}}$, respectively.
In comparison, the experimental values at room temperature are 0.23,
0.48, and 0.93~Å$^2$/ps \cite{holz2000temperature, johnaelliot1990estimation, buxton1988critical}, respectively, and these diffusion coefficients are expected to be even higher at 350~K.
This, by itself, leads to an underestimation of the diffusion contribution in $k_2$, so the overestimation of the overall $k_2$ should stem from the reaction-controlled contribution, which is also ultimately determined by the underlying density functional.
Another remark is that our simulation is run with NVT at 1~g/cm$^3$,
while the experimental density slightly decreases with temperature. However,
within our temperature range, this should be a relatively small effect. An additional source of error comes from neglected nuclear quantum effects, which would presumably increase $k_2$ due to a slightly delocalized proton. 
Nevertheless, the ratio of the diffusion coefficients between H$_2$O, e$^-_{\mathrm{(aq)}}$, and H$^+_{\mathrm{(aq)}}$ in our model
is in reasonable agreement with experiment, so the qualitative picture, as well as the temperature dependence, should remain physically consistent.

\emph{Equilibrium constant.}
In addition to $k_2$, Reaction~(\ref{reaction}) is also characterized by its equilibrium constant, denoted $K_1$.
This can be calculated from the expression \cite{davies2002estimating, calegari2023probing}
\begin{equation}\label{K}
    \mathrm{p}K_1 = \log_{10} c_0 \int_0^{r_c} e^{-\beta (W(r) - W(r_c))} 4\pi r^2 dr,
\end{equation}
which relates $K_1$ to the potential of mean force $W(r)$,
a function of the distance $r$ between the excess proton H$^+$ and the electron center $\mathbf{c}_{\mathrm{e}^-}$.
When $r$ is around zero, it represents the state of H$\cdot_{\mathrm{(aq)}}$. Beyond a certain distance $r_c$, $W(r)$ plateaus and $W(r_c) = W(\infty)$, representing independent e$^-_{\mathrm{(aq)}}$ and H$^+_{\mathrm{(aq)}}$.
$c_0 = 1\text{ M}$ is the standard concentration.

Enhanced sampling is required to determine $W(r)$, as the backward reaction is not spontaneous.
Here, we use the value of $V_{\mathrm{emb}}$ at the position of H$^+$ as our collective variable (CV).
This reflects the electrostatics of the solvent and is well correlated with the progress of the reaction, 
where $V_{\mathrm{emb}}(\mathrm{H}^+)$ is low in the product state H$\cdot_{\mathrm{(aq)}}$ and increases as H$^+$ and e$^-$ start to separate.
We add a bias potential to raise the free energy corresponding to H$\cdot_{\mathrm{(aq)}}$ \footnote{This suffices for our equilibrium sampling. It
avoids differentiating $\mathbf{c}_{\mathrm{e}^-}$ through the iterative eigen-solver and avoids the discontinuity as H$^+$ hops, which would arise with a common enhanced sampling algorithm with $r$ as the CV.},
making the reaction reversible within our simulation time.
Then, from the biased molecular dynamics simulation,
$W(r)$ and~(\ref{K}) can be evaluated using the properly weighted histogram of the trajectory.

We perform the calculation at 350, 400, and 450~K with up to 512 H$_2$O molecules plus one proton and one electron.

We find an acceptable convergence of $W(r)$ for the 512~H$_2$O size, with $r_c$ set to 10~Å.
A van~'t~Hoff plot of p$K_1$ against inverse temperature is shown in Fig.~\ref{fig:rate}(b).
Our p$K_1$ decreases from 9.3 at 350~K to 7.3 at 450~K;
the positive slope of p$K_1$ vs $T^{-1}$ indicates an endothermic reaction, and the magnitude of the slope agrees well with experiment \cite{shiraishi1994temperature}.
We also observe a small positive curvature consistent with experimental trends,
which has been proposed to be related to a decrease in the dielectric constant of water at higher temperatures \cite{shiraishi1994temperature}.
At 350~K, p$K_1 = 9.3$ corresponds to a reaction free energy of 0.65~eV,
which is overestimated by 0.07~eV compared to experiment.
We also remark that the same reaction in vacuum,
the ionization of a hydrogen atom, takes an energy of 13.6~eV.
This underscores the dramatic stabilization effect provided by the solvent environment.

\emph{Conclusion.}
We have developed an electron-aware machine-learning method to model e$^-_{\mathrm{(aq)}}$ and its reaction with H$^+_{\mathrm{(aq)}}$,
relying on a physics-motivated electrostatic embedding to solve for the excess electron. The model is accurate, robust, efficient, and extensible.
It exhibits a proton transfer mechanism,
and we determine the reaction rate and equilibrium constant over a temperature range,
with results in agreement with experiments.
Assessment of more accurate electronic structure methods, nuclear quantum effects, and the detailed analysis of the transition state are left for future work.
In addition, we expect our method to naturally apply to a broad class of solvated-electron reactions.
It would also be interesting to extend the framework
to nonadiabatic electron transfer processes, other small polarons, bound excitons, interfacial systems \cite{ding2022mgo}, and situations where explicit electron modeling is important.

\emph{Data availability.}
The data and the trained model used in this work are available at Zenodo \cite{gao2025-dataset}. The code for DPLR-q is available on GitHub \cite{dplrq-code}. 

\emph{Acknowledgements.} 
R.G. and R.C. were supported by the U.S. Department of Energy (DOE) through the project Multiscale Chemistry and Physics in Materials and Fluids (Award No. DE-SC0026745) and the Computational Chemical Science Center: Chemistry in Solution and at Interfaces (Award No. DE-SC0019394), and by a gift from Seven Research.
R.G. acknowledges the hospitality of Lawrence Berkeley National Laboratory, where a portion of this research was performed during a research visit. P.X. was supported by Luis W. Alvarez Fellowship and the DOE Advanced Scientific Computing Research (ASCR) Applied Mathematics program under Contract No. DE-AC02-05CH11231. 
This research used resources of the National Energy Research
Scientific Computing Center (NERSC), a U.S. Department of Energy
Office of Science User Facility operated under Contract No.
DE-AC02-05CH11231 and using NERSC award BES-ERCAP0037271.

\bibliographystyle{apsrev4-2}
\bibliography{reference}
\clearpage
\appendix*
\section{End Matter}

\emph{Appendix A. Details on DFT calculations and model training.}
We run all DFT calculations using the hybrid functional PBEh($\alpha$) plus the rVV10 van der Waals correction.
The fraction of Fock exchange $\alpha$ is set to 0.4, and the parameter $b$ in rVV10 is set to 5.3, as proposed in Ref.~\cite{ambrosio2016structural}.
We use the Goedecker–Teter–Hutter pseudopotentials \cite{goedecker1996separable}, 
a triple-$\zeta$ polarized basis \cite{vandevondele2007gaussian}, an 800~Ry cutoff for the expansion of density in plane waves, and the auxiliary density matrix method with the cFIT3 basis set \cite{guidon2010auxiliary}.
We perform a spin-polarized calculation, as implemented in the CP2K package \cite{kuhne2020cp2k}. 

To generate an initial DFT dataset, we start with a periodic box of 64~H$_2$O molecules at 1 g/cm$^3$ in an NVT ensemble at 350~K.
A proton is added and equilibrated for 200~fs. We then add an excess e$^-$, whose initial ground state is delocalized. Within 200--400~fs, e$^-$ creates a cavity to localize itself, forming e$^-_{\mathrm{(aq)}}$ \cite{pizzochero2019picture}. We prepare 15 such initial configurations and run DFT trajectories in parallel for 10--20 ps each; a few of them end up reacting within this timescale.
From these trajectories, we extract $3\times10^{4}$ configurations covering both reactant and product states as our initial training data.

\begin{figure*}[t]
  \centering
  
  \begin{minipage}[b]{0.355\linewidth}
    \begin{overpic}[width=\linewidth]{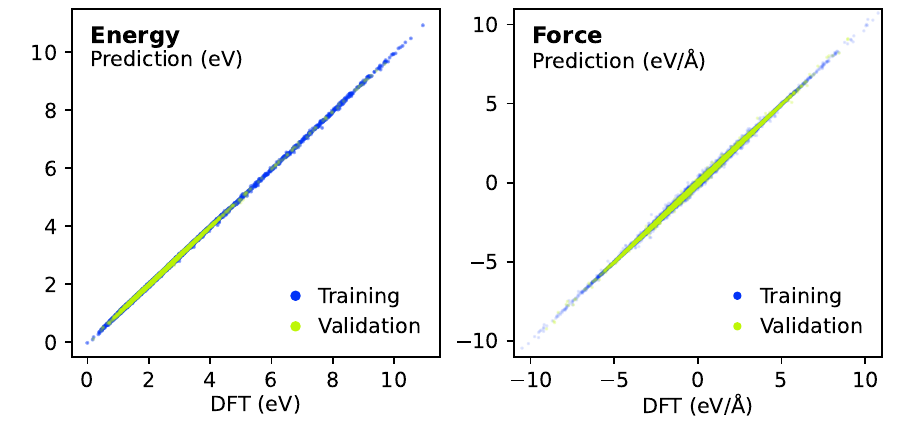}
      \put(0,47){(a)} 
    \end{overpic}
  \end{minipage}
  \hspace{1pt} 
  \begin{minipage}[b]{0.225\linewidth}
    \begin{overpic}[width=\linewidth, trim=12 10 8 10, clip]{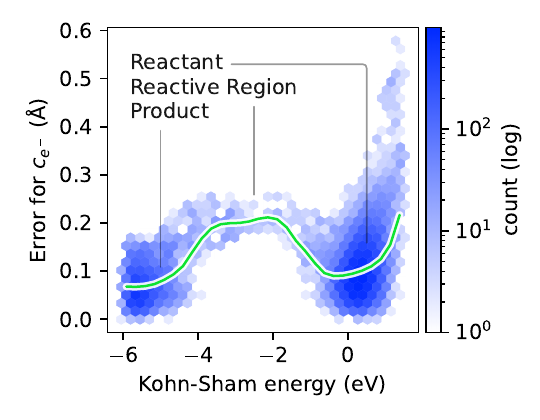}
      \put(-5,72){(b)}
    \end{overpic}
  \end{minipage}
  \hspace{1pt} 
  \begin{minipage}[b]{0.38\linewidth}
    \begin{overpic}[width=\linewidth, trim=15 0 0 0, clip]{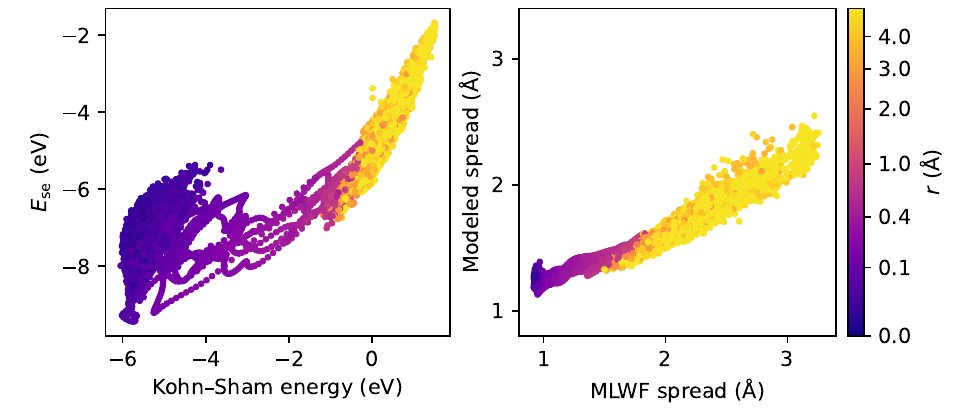}
      \put(-3,43){(c)}
    \end{overpic}
  \end{minipage}

  \vspace{-2mm}
  \caption{
  (a) Energy/force parity plots: DFT vs trained DPLR-q.  
  (b) $\mathbf{c}_{e^-}$–MLWF center discrepancy vs Kohn–Sham e$^-$ energy; density (blue) and binned mean error (green). 
  (c) Modeled vs DFT e$^-$ energy/spread; color indicates the distance $r$ between H$^+$ and e$^-$.
  Panel (a) uses the full dataset; (b,c) use configurations from the initial DFT trajectories.
  }
  \label{fig:combined_results}
\end{figure*}

We first train an initial model on this dataset. It enables the active learning process, where we run simulations with our model and sample additional configurations using a model deviation metric~\cite{zhang2020dp}. In total, about 1,300 additional configurations are added to the dataset. The final dataset contains 31,518 configurations of 64 water molecules plus one proton and one electron, which we use to train the production model. Validation errors are reported using a 90\%--10\% random train--validation split of the final dataset. Figure~\ref{fig:combined_results}(a) compares model-predicted total energies and forces with DFT values and demonstrates the excellent accuracy and generalization of our model. Our production model can run at 3.3 ns/day for the 512-H$_2$O cell on an NVIDIA A100 GPU.

\emph{Appendix B. Selection of the Gaussian smoothing parameter.}
In the Gaussian smoothing of $\rho_{\textrm{emb}}$ using the kernel $e^{-(\beta r)^2}$, we find that a smaller $\beta$ (stronger smoothing) increases the deviation of the modeled center $\mathbf{c}_{e^-}$ from the MLWF center, whereas a larger $\beta$ requires a denser grid for convergence. 
We adopt $\beta = 2$~Å$^{-1}$ with a grid spacing of about 0.26~Å, which balances these factors.
The resulting average distance between $\mathbf{c}_{e^-}$ and the MLWF center is $0.10$~Å \cite{dwc}.
The two terms $E_{\mathrm{se}}$ and $E_{\mathrm{lr}}$ partially cancel due to the screening of water around e$^-_{\mathrm{(aq)}}$ or the proton's charge in H$\cdot_{\mathrm{(aq)}}$.
A different smoothing parameter $\beta_{\mathrm{lr}} = $ 0.6~Å$^{-1}$ is used in the calculation of $E_{\mathrm{lr}}$ to numerically minimize $|\nabla_{\mathbf{R}}(E_{\mathrm{se}} + E_{\mathrm{lr}})|$ and thus enhance this cancellation. Eventually,
$\nabla_{\mathbf{R}} E_{\mathrm{sr}}$ remains the primary contributor to the forces.
The precise values of $\beta$ and $\beta_{\mathrm{lr}}$ are not critical within a reasonable range, since any differences would be absorbed into $E_{\mathrm{sr}}$.

\emph{Appendix C. Investigating the solution of the auxiliary Schrödinger equation.}
The error for the center of mass $\mathbf{c}_{\mathrm{e}^-}$ obtained from solving Eq.~(\ref{se}) is further analyzed in Figure~\ref{fig:combined_results}(b) across the reactant, product, and reactive regions. It can be seen that the error is relatively larger ($0.2$ \AA) compared to the average error ($0.1$ \AA) in the reactive region and in reactant configurations with a relatively more delocalized e$^-$ (corresponding to the right end). These are cases when the approximation made in Eq.~(\ref{se}) is slightly less accurate. Nevertheless, as we have quantified in footnote~\cite{dwc}, this level of error should not affect the overall accuracy of the final ML potential.

In Fig.~\ref{fig:combined_results}(c), we plot the modeled electron's energy $E_{\mathrm{se}}$ versus the Kohn-Sham energy level of e$^-$, and the modeled electron's spread versus the MLWF spread of e$^-$ from DFT.
For an isolated e$^-_{\mathrm{(aq)}}$,
the Kohn-Sham energy level lies well within the band gap,
typically about 2--3~eV below the conduction band minimum,
and the MLWF spread fluctuates around 2~Å.
During the reaction, the Kohn-Sham energy level decreases to just slightly above the valence band maximum, and the MLWF spread drops to just under 1~Å.
The modeled energy and spread are correlated with the DFT counterparts but are not equal.
The range of energy variation is similar, 
and the spread variation for $n_{\mathrm{e}^-}$ is somewhat smaller than that in DFT.
This behavior likely reflects the absence of self-interaction in our one-electron Schrödinger equation~(\ref{se}),
whereas DFT, even at the hybrid level, still retains some residual self-interaction.
Again, the Kohn-Sham energy and wavefunction are not our targets, and the comparison here is only a sanity check. As long as $\mathbf{c}_{\mathrm{e}^-}$ agrees with the MLWF center,
these differences are short-ranged and are absorbed by the neural network in $E_{\mathrm{sr}}$.

\begin{figure}[h]
  \centering
  \begin{overpic}[width=0.5\columnwidth]{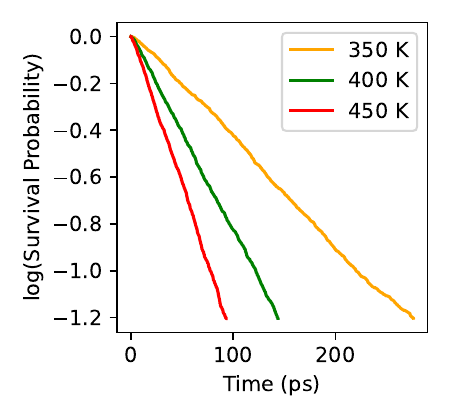}
    \put(3,90){\footnotesize (a)}
  \end{overpic}\hfill
  \begin{overpic}[width=0.5\columnwidth]{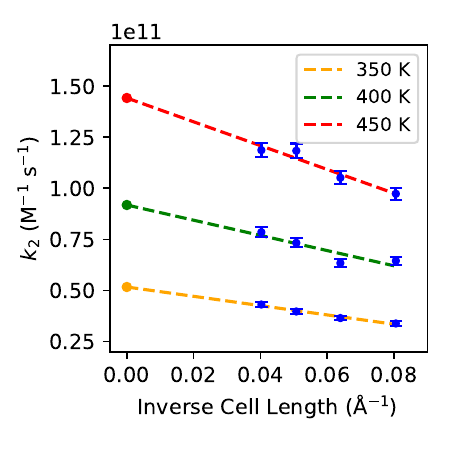}
    \put(3,90){\footnotesize (b)}
  \end{overpic}
  \caption{(a) The obtained $\ln(\hat{s}(t))$ using a 512-H$_2$O system and 2000 independent trajectories.
           (b) Extrapolation of $k_2$ to the dilute limit.}
  \label{fig:rate_analysis}
\end{figure}

\emph{Appendix D. Details of the reaction-rate calculation.}
An example of the obtained log-survival probability versus time is shown in Fig.~\ref{fig:rate_analysis}(a).
One can see a clear linear behavior, where the slope is equal to $\lambda$.
We remark that if the initial configurations are not sampled from the QSD,
the log-survival probability would be curved, and the estimation of $\lambda$ would be biased.
In Fig.~\ref{fig:rate_analysis}(b), we show the extrapolation of the rate constant $k_2$ to the dilute limit.
We fit the results from the calculations of 64, 128, 256, and 512-H$_2$O molecules.
From our numerical results, the finite-size error scales approximately as $1/L$,
as expected from the 
hydrodynamic finite-size correction to the diffusion constant under
periodic boundary conditions~\cite{yeh2004system} 
and the leading $1/L$ Ewald electrostatic self-interaction term for charged periodic systems~\cite{MakovPayne1995}.

\begin{figure}[h]
  \centering
  \begin{overpic}[width=0.53\columnwidth]{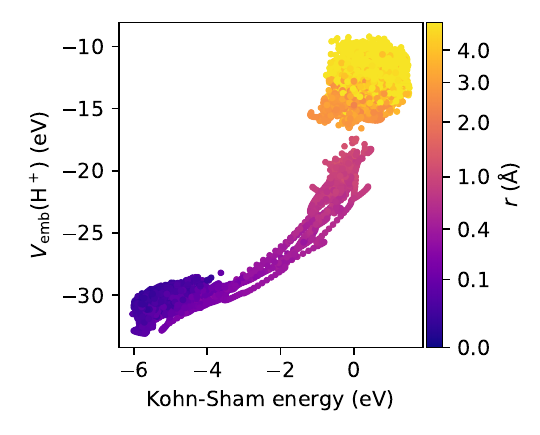}
    \put(2,72){\footnotesize (a)}
  \end{overpic}\hfill
  \begin{overpic}[width=0.46\columnwidth]{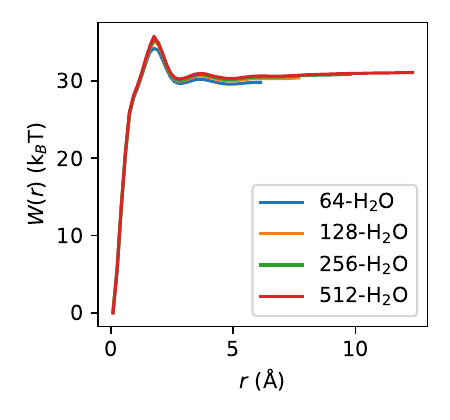}
    \put(3,83){\footnotesize (b)}
  \end{overpic}
  \caption{(a) $V_{\mathrm{emb}}(\mathrm{H}^+)$ vs DFT energy of e$^-$. The distance $r$ between H$^+$ and e$^-$ is indicated by color.
           (b) Potential of mean force $W(r)$ at 400~K for different system sizes.}
  \label{fig:wr_analysis}
\end{figure}

\emph{Appendix E. Details of the equilibrium-constant calculation.}
We first plot $V_{\mathrm{emb}}(\mathrm{H}^+)$ versus the Kohn-Sham energy of e$^-$ in Fig.~\ref{fig:wr_analysis}(a) using the data from DFT trajectories.
The upper right blob corresponds to cases where the identified proton is not a neighbor of e$^-$.
The lower branch is the reaction pathway of interest, with the two quantities tightly correlated. This suggests that $V_{\mathrm{emb}}(\mathrm{H}^+)$ can serve as a CV for equilibrium sampling.
The product H$\cdot_{\mathrm{(aq)}}$ corresponds to $V_{\mathrm{emb}}\!\!\approx\!\!-30$~eV.
We add a bias that lifts the region below $\sim\!-25$ eV to sample both reaction directions.
The precise value of $V_{\mathrm{emb}}(\mathrm{H}^+)$ is not physically relevant, because it is $\beta$-dependent.

In Fig.~\ref{fig:wr_analysis}(b), we plot the obtained potential of mean force $W(r)$ for different system sizes at 400~K as an example.
Our sampled trajectory is reweighted to the Boltzmann distribution according to the bias potential, and $W(r)$ is obtained from the logarithm of the sample histogram. The curves are normalized such that $W(0) = 0$, and we plot up to half of the box length.
We see acceptable convergence at the 512-H$_2$O size.
The Coulombic attraction between e$^-_{\mathrm{(aq)}}$ and H$^+_{\mathrm{(aq)}}$ is well screened by water,
leaving a vanishingly small tail in $W(r)$ at large $r$.
The peak around 2~Å is due to the discontinuous change in the identity of H$^+$; thus,
 the histogram contains few samples in this region, and the resulting free energy exhibits a peak.
Nevertheless, the evaluation of p$K_1$ through the integral in (\ref{K}) remains well-defined and unaffected by such discontinuities.

\emph{Appendix F. Comments on the failure of a standard neural network potential.} It is difficult for a neural network potential alone to learn the reaction process, although this is not obvious a priori.
Even if the electron is not represented,
it may still be implicitly inferred from the modified atomic environment surrounding the cavity.
One may thus believe that this would, in principle, allow an ML force field to correctly propagate the dynamics of the system. 
However, when we train such a force field and run molecular dynamics simulations,
we find that it is at risk of occasionally losing track of the local structure around e$^-_{\mathrm{(aq)}}$,
causing the configuration to collapse to pure water.
Once this happens, it is difficult for the system to recover the e$^-_{\mathrm{(aq)}}$ state,
and even if it does, such dynamics cannot be reproduced by DFT and should not have occurred in the first place.
Attempts have been made for a system of e$^-_{\mathrm{(aq)}}$ in water \cite{gao2024enhanced}, where such failures are mitigated by leveraging a more powerful neural network.
However, we find that for modeling reaction~(\ref{reaction}), where the H$^+$ further complicates the system, it remains difficult to eliminate such failures on nanosecond timescales. In our attempts, there is little sign of improvement when we further enhance the neural network or add more training data with active learning.
We also note that in Refs.~\cite{lan2021simulating, lan2022temperature}, ML models for e$^-_{\mathrm{(aq)}}$ have been trained; however, the former is limited to a small simulation cell and is not extensible to larger systems, while the latter is not a standalone ML force field and requires DFT calculations to aid the simulation. 
Overall, we suggest that the failure of standard finite-range neural networks stems from their inability to capture the long-tailed Coulombic interactions.
We take the view that a more physical treatment of e$^-$, where the electrostatics responsible for its localization is explicitly modeled, offers a more reliable route than relying on purely data-driven neural-network fitting.

\end{document}